\documentclass[11pt]{article}
\usepackage{amsmath,amssymb,latexsym}
\usepackage{graphicx}

\newcommand\rf[1]{(\ref{eq:#1})}
\newcommand\lab[1]{\label{eq:#1}}
\newcommand\nonu{\nonumber}
\newcommand\br{\begin{eqnarray}}
\newcommand\er{\end{eqnarray}}
\newcommand\be{\begin{equation}}
\newcommand\ee{\end{equation}}

\newcommand\foot[1]{\footnotemark\footnotetext{#1}}

\newcommand\llb{\left\lbrack}
\newcommand\rrb{\right\rbrack}

\newcommand\lcurl{\left\{}
\newcommand\rcurl{\right\}}
\renewcommand\({\left(}
\renewcommand\){\right)}
\newcommand\bv{\bigm\vert}               

\newcommand\bc{\begin{center}}
\newcommand\ec{\end{center}}

\newcommand\partder[2]{\frac{{\partial {#1}}}{{\partial {#2}}}}

\renewcommand\a{\alpha}
\renewcommand\b{\beta}

\newcommand\D{\Delta}

\newcommand\vareps{\varepsilon}
\newcommand\g{\gamma}
\newcommand\G{\Gamma}

\newcommand\h{\frac{1}{2}}
\renewcommand\k{\kappa}
\renewcommand\l{\lambda}

\newcommand\m{\mu}
\newcommand\n{\nu}

\newcommand\om{\omega}

\newcommand\vp{\varphi}
\renewcommand\P{\Phi}
\newcommand\pa{\partial}

\newcommand\pr{\prime}

\newcommand\s{\sigma}
\renewcommand\S{\Sigma}
\renewcommand\t{\tau}
\renewcommand\th{\theta}

\newcommand\wti{\widetilde}

\newcommand\cA{{\mathcal A}}

\newcommand\cE{{\mathcal E}}
\newcommand\cF{{\mathcal F}}

\newcommand{\ct}[1]{\cite{#1}}
\newcommand{\bib}[1]{\bibitem{#1}}

\newcommand\NPB[3]{\textsl{Nucl. Phys.} \textbf{B#1} (#2) #3}

\newcommand\PRD[3]{\textsl{Phys. Rev.} \textbf{D#1} (#2) #3}

\newcommand\CQG[3]{\textsl{Class. Quantum Grav.} \textbf{#1} (#2) #3}

\newcommand\IJMPA[3]{\textsl{Int. J. Mod. Phys.} \textbf{A#1} (#2) #3}

\newcommand\Xdot{\stackrel{.}{X}}
\newcommand\ydot{\stackrel{.}{y}}
\newcommand\yddot{\stackrel{..}{y}}
\newcommand\rdot{\stackrel{.}{r}}

\newcommand\vpdot{\stackrel{.}{\varphi}}

\begin{document}

\title{Lightlike $p$-Branes: Mass ``Inflation'' \\
and Lightlike Braneworlds}

\author{Eduardo Guendelman and Alexander Kaganovich\\
\small\it Department of Physics, Ben-Gurion University of the Negev,\\[-1.mm]
\small\it P.O.Box 653, IL-84105 ~Beer-Sheva, Israel  \\[-1.mm]
\small\it guendel@bbgu.ac.il , alexk@bgu.ac.il\\
{} \\
Emil Nissimov and Svetlana Pacheva\\
\small\it Institute for Nuclear Research and Nuclear Energy,\\[-1.mm]
\small\it Bulgarian Academy of Sciences, Sofia, Bulgaria  \\[-1.mm]
\small\it email: nissimov@inrne.bas.bg, svetlana@inrne.bas.bg}

\maketitle

\begin{abstract}\foot{To appear in {\sl ``Fifth Summer School in 
Modern Mathematical Physics''}, B. Dragovich and Z. Rakic (eds.), 
Belgrade Inst. Phys. Press, 2009.}
Lagrangian description of lightlike $p$-branes is presented in two equivalent forms
-- a Polyakov-type formulation and a dual to it Nambu-Goto-type formulation. 
Next, the properties of lightlike brane dynamics in generic gravitational backgrounds
of spherically symmetric and axially symmetric type are discussed in some
detail: ``horizon straddling'' and ``mass inflation'' effects for codimension-one 
lightlike branes and ground state behavior of codimension-two lightlike 
``braneworlds''. 

\end{abstract}

\section{Introduction}
Lightlike branes (\textsl{LL-branes}, for short) are of particular interest in 
general relativity primarily due to their role in the
effective treatment of many cosmological and astrophysical effects:
(i) impulsive lightlike signals arising in cataclysmic astrophysical events
\ct{barrabes-hogan}; 
(ii) the ``membrane paradigm'' theory of black hole physics 
\ct{membrane-paradigm}; 
(iii) thin-wall description of domain walls coupled to gravity
\ct{Israel-66,Barrabes-Israel-Hooft}.
More recently \textsl{LL-branes} became significant also in the context of
modern non-perturbative string theory \ct{nonperturb-string}.

Here we first present explicit reparametrization invariant 
$(p+1)$-dimen\-sio\-nal
world-volume actions describing \textsl{LL-brane} dynamics in two equivalent 
forms: (i) Polyakov-type formulation, and (ii) Nambu-Goto-type formulation 
dual to the first one. 

Unlike ordinary Nambu-Goto $p$-branes (describing massive brane modes) our models 
yield intrinsically lightlike $p$-branes (the induced metric becoming singular 
on-shell) with the additional crucial property of the {\em brane tension} appearing 
as a {\em non-trivial dynamical degree of freedom}. The latter characteristic
feature significantly distinguishes our lightlike $p$-brane models from the
previously proposed {\em tensionless} $p$-branes (for a review, see
\textsl{e.g.} \ct{tensionless}) which rather resemble a $p$-dimensional 
continuous distribution of massless point-particles.

Next we discuss the properties of \textsl{LL-brane} dynamics in generic 
gravitational backgrounds. The case with two extra dimensions (codimension-two 
\textsl{LL-branes}) is studied from the point of view of ``braneworld'' scenarios.
Unlike conventional braneworlds, where the underlying branes are of Nambu-Goto type 
and in their ground state they position themselves at some
fixed point in the extra dimensions of the bulk space-time, our lightlike
braneworlds perform in the ground state non-trivial motions in the extra
dimensions -- planar circular, spiral winding \textsl{etc} depending on the
topology of the extra dimensions.

The special case of codimension-one \textsl{LL-branes} is qualitatively
different. Here the \textsl{LL-brane} dynamics dictates 
that the bulk space-time with a bulk metric of spherically or axially symmetric type
must possess an event horizon which is automatically occupied by the 
\textsl{LL-brane} (``horizon straddling'').
We study several cases of ``horizon straddling'' solutions. In the case of
Kerr ``horizon straddling'' by a \textsl{LL-brane} there is the additional effect 
of brane rotation ``dragged'' by the Kerr black hole.

For the inner Reissner-Nordstr{\"o}m horizon we find a time symmetric ``mass
inflation'' effect, which also holds for de Sitter horizon. In this case
the dynamical tension of the \textsl{LL-brane} blows up as time approaches
$\pm \infty$ due to its exponential quadratic time dependence.
For the Schwarzschild and the outer Reissner-Nordstr{\"o}m horizons, on the
other hand, we obtain ``mass deflationary'' scenarios where the dynamical 
\textsl{LL-brane} tension vanishes at large positive or large negative times.
Another set of solutions with asymmetric (w.r.t. $t \to -t$) exponential
linear time dependence of the \textsl{LL-brane} tension also exists. 
By fine tuning one can obtain a constant time-independent brane tension 
as a special case. The latter holds in particular for \textsl{LL-branes}
moving in extremal Reissner-Nordstr{\"o}m or maximally rotating Kerr black
hole backgrounds.

\section{World-Volume Actions for Lightlike Branes}
In refs.\ct{will-brane-all,varna-07,inflation-07} we have proposed the
following generalized Polyakov-type formulation of the Lagrangian dynamics of
\textsl{LL-branes} in terms of the world-volume action:
\br
S = \int d^{p+1}\s \,\P (\vp) \llb - \h \g^{ab} g_{ab} + L\!\( F^2\)\rrb \; .
\lab{LL-brane}
\er
Here $\g_{ab}$ denotes the intrinsic Riemannian metric on the world-volume;
\be
g_{ab} \equiv \pa_a X^{\m} \pa_b X^{\n} G_{\m\n}(X)
\lab{ind-metric}
\ee
is the induced metric (the latter becomes {\em singular} on-shell -- lightlikeness, 
cf. second Eq.\rf{phi-gamma-eqs} below);
\be
\P (\vp) \equiv \frac{1}{(p+1)!} \vareps_{I_1\ldots I_{p+1}}
\vareps^{a_1\ldots a_{p+1}} \pa_{a_1} \vp^{I_1}\ldots \pa_{a_{p+1}} \vp^{I_{p+1}}
\lab{mod-measure-p}
\ee
is an alternative non-Riemannian reparametrization-covariant integration measure 
density replacing the standard $\sqrt{-\g} \equiv \sqrt{-\det \Vert \g_{ab}\Vert}$
and built from auxiliary world-volume scalars $\lcurl \vp^I \rcurl_{I=1}^{p+1}$;
\be
F_{a_1 \ldots a_{p}} = p \pa_{[a_1} A_{a_2\ldots a_{p}]} \quad ,\quad
F^{\ast a} = \frac{1}{p!} \frac{\vareps^{a a_1\ldots a_p}}{\sqrt{-\g}}
F_{a_1 \ldots a_{p}}
\lab{p-rank}
\ee
are the field-strength and its dual one of an auxiliary world-volume $(p-1)$-rank 
antisymmetric tensor gauge field $A_{a_1\ldots a_{p-1}}$ with Lagrangian $L(F^2)$ 
~($F^2 \equiv F_{a_1 \ldots a_{p}} F_{b_1 \ldots b_{p}} 
\g^{a_1 b_1} \ldots \g^{a_p b_p}$).

Equivalently one can rewrite \rf{LL-brane} as:
\br
S = \int d^{p+1}\!\!\s \,\chi \sqrt{-\g} 
\llb -\h \g^{ab} g_{ab} + L\!\( F^2\)\rrb \quad, \;\; 
\chi \equiv \frac{\P (\vp)}{\sqrt{-\g}}
\lab{LL-brane-chi}
\er
The composite field $\chi$ plays the role of a 
{\em dynamical (variable) brane tension}.

For the special choice $L\!\( F^2\)= \( F^2\)^{1/p}$ the above action  
becomes invariant under Weyl (conformal) symmetry: 
\br
\g_{ab} \longrightarrow \g^{\pr}_{ab} = \rho\,\g_{ab}  \quad ,\quad
\vp^{i} \longrightarrow \vp^{\pr\, i} = \vp^{\pr\, i} (\vp)
\lab{Weyl-conf}
\er
with Jacobian $\det \Bigl\Vert \frac{\pa\vp^{\pr\, i}}{\pa\vp^j} \Bigr\Vert = \rho$. 

Consider now the equations of motion corresponding to \rf{LL-brane} 
w.r.t. $\vp^I$ and $\g^{ab}$:
\br
\h \g^{cd} g_{cd} - L(F^2) = M \quad , \quad
\h g_{ab} - F^2 L^{\pr}(F^2) \llb\g_{ab} 
- \frac{F^{*}_a F^{*}_b}{F^{*\, 2}}\rrb = 0  \; .
\lab{phi-gamma-eqs}
\er
Here $M$ is an integration constant and $F^{*\, a}$ is the dual field
strength \rf{p-rank}. Both Eqs.\rf{phi-gamma-eqs} imply the constraint
$L\!\( F^2\) - p F^2 L^\pr\!\( F^2\) + M = 0$, \textsl{i.e.}
\br
F^2 = F^2 (M) = \mathrm{const} ~\mathrm{on-shell} \; .
\lab{F2-const}
\er
The second Eq.\rf{phi-gamma-eqs} exhibits {\em on-shell singularity} 
of the induced metric \rf{ind-metric}: 
\br
g_{ab}F^{*\, b}=0 \; .
\lab{on-shell-singular}
\er

Further, the equations of motion w.r.t. world-volume gauge field 
$A_{a_1\ldots a_{p-1}}$ (with $\chi$ as defined in \rf{LL-brane-chi} and
accounting for the constraint \rf{F2-const}):
\br
\pa_{[a}\( F^{\ast}_{b]}\, \chi\) = 0 
\lab{A-eqs}
\er
allow us to introduce the dual ``gauge'' potential $u$:
\be
F^{\ast}_{a} = \mathrm{const}\, \frac{1}{\chi} \pa_a u \; .
\lab{u-def}
\ee
We can rewrite second Eq.\rf{phi-gamma-eqs} (the lightlike constraint)
in terms of the dual potential $u$ as:
\be
\g_{ab} = \frac{1}{2a_0}\, g_{ab} - \frac{2}{\chi^2}\,\pa_a u \pa_b u \quad ,
\quad a_0 \equiv F^2 L^{\pr}\( F^2\)\bv_{F^2=F^2(M)} = \mathrm{const}
\lab{gamma-eqs-u}
\ee
($L^\pr(F^2)$ denotes derivative of $L(F^2)$ w.r.t. the argument $F^2$).
From \rf{u-def} and \rf{F2-const} we have the relation: 
\be
\chi^2 = -2 \g^{ab} \pa_a u \pa_b u \; ,
\lab{chi2-eq}
\ee
and the Bianchi identity $\pa_a F^{\ast\, a}=0$ becomes:
\be
\pa_a \Bigl( \frac{1}{\chi}\sqrt{-\g} \g^{ab}\pa_b u\Bigr) = 0  \; .
\lab{Bianchi-id}
\ee

Finally, the $X^\m$ equations of motion produced by the \rf{LL-brane} read:
\br
\pa_a \(\chi \sqrt{-\g} \g^{ab} \pa_b X^\m\) + 
\chi \sqrt{-\g} \g^{ab} \pa_a X^\n \pa_b X^\l \G^\m_{\n\l}(X) = 0  \;
\lab{X-eqs}
\er
where $\G^\m_{\n\l}=\h G^{\m\k}\(\pa_\n G_{\k\l}+\pa_\l G_{\k\n}-\pa_\k G_{\n\l}\)$
is the Christoffel connection for the external metric.

It is now straightforward to prove that the system of equations 
\rf{chi2-eq}--\rf{X-eqs} for $\( X^\m,u,\chi\)$, which are equivalent to the 
equations of motion \rf{phi-gamma-eqs}--\rf{A-eqs},\rf{X-eqs} resulting from the 
original Polyakov-type \textsl{LL-brane} action \rf{LL-brane}, can be equivalently 
derived from the following {\em dual} Nambu-Goto-type world-volume action: 
\br
S_{\rm NG} = - \int d^{p+1}\s \, T 
\sqrt{- \det\Vert g_{ab} - \frac{1}{T^2}\pa_a u \pa_b u\Vert}  \; .
\lab{LL-action-NG}
\er
Here $g_{ab}$ is the induced metric \rf{ind-metric};
$T$ is {\em dynamical} tension simply related to the dynamical tension 
$\chi$ from the Polyakov-type formulation \rf{LL-brane-chi} as
$T^2= \frac{\chi^2}{4a_0}$ with $a_0$ -- same constant as in \rf{gamma-eqs-u}.

Henceforth we will consider the initial Polyakov-type form \rf{LL-brane} of the
\textsl{LL-brane} world-volume action. Invariance under world-volume 
reparametrizations allows to introduce the standard synchronous gauge-fixing 
conditions:
\be
\g^{0i} = 0 \;\; (i=1,\ldots,p) \; ,\; \g^{00} = -1
\lab{gauge-fix}
\ee
Also, in what follows we will use a natural ansatz for the ``electric'' part of the 
auxiliary world-volume gauge field-strength:
\be
F^{\ast i}= 0 \;\; (i=1,{\ldots},p) \quad ,\quad \mathrm{i.e.} \;\;
F_{0 i_1 \ldots i_{p-1}} = 0 \; ,
\lab{F-ansatz}
\ee
The Bianchi identity ($\pa_a F^{\ast\, a}=0$) together with 
\rf{gauge-fix}--\rf{F-ansatz} and the definition for the dual field-strength
in \rf{p-rank} imply:
\be
\pa_0 \g^{(p)} = 0 \quad \mathrm{where}\;\; \g^{(p)} \equiv \det\Vert\g_{ij}\Vert \; .
\lab{gamma-p-0}
\ee
Then \textsl{LL-brane} equations of motion acquire the form 
(recall definition of $g_{ab}$ \rf{ind-metric}):
\be
g_{00}\equiv \Xdot^\m\!\! G_{\m\n}\!\! \Xdot^\n = 0 \quad ,\quad g_{0i} = 0 \quad ,\quad
g_{ij} - 2a_0\, \g_{ij} = 0
\lab{gamma-eqs-0}
\ee
(Virasoro-like constraints), where the $M$-dependent constant $a_0$ (the
same as in \rf{gamma-eqs-u}) must be strictly positive;
\be
\pa_i \chi = 0 \qquad (\mathrm{remnant ~of ~Eq.\rf{A-eqs}})\; ;
\lab{A-eqs-0}
\ee

\vspace{-0.6cm}
\br
-\sqrt{\g^{(p)}} \pa_0 \(\chi \pa_0 X^\m\) +
\pa_i\(\chi\sqrt{\g^{(p)}} \g^{ij} \pa_j X^\m\)
\nonu \\
+ \chi\sqrt{\g^{(p)}} \(-\pa_0 X^\n \pa_0 X^\l + \g^{kl} \pa_k X^\n \pa_l X^\l\)
\G^\m_{\n\l} = 0 \; .
\lab{X-eqs-0}
\er

\section{Lightlike Branes in Gravitational Backgrounds:\\ Codimension-Two}
Let us split the bulk space-time coordinates as:
\br
\( X^\m\) = \( x^a, y^\a\) \equiv \( x^0 \equiv t, x^i, y^\a\)
\lab{coord-split} \\
a=0,1,\ldots, p \;\; ,\;\; i=1,\ldots, p \;\; ,\;\;
\a = 1,\ldots, D-(p+1)
\nonu
\er
and consider background metrics $G_{\m\n}$ of the form:
\be
ds^2 = - A(t,y)(dt)^2 + C(t,y) h_{ij}(\vec{x}) dx^i dx^j + 
B_{\a\b}(t,y) dy^\a dy^\b
\lab{nonstatic-metric}
\ee

Here we will discuss the simplest non-trivial ansatz for the \textsl{LL-brane}
embedding coordinates:
\be
X^a \equiv x^a = \s^a \quad, \quad X^{p+\a} \equiv y^\a = y^\a (\t) \quad,
\quad \t \equiv \s^0  \; .
\lab{X-ansatz}
\ee
and will take the particular solution $\chi = \mathrm{const}$ of Eq.\rf{A-eqs-0}
for the dynamical tension (for more general time-dependent dynamical tension 
solutions, see next Section). Then 

Now the \textsl{LL-brane} (gauge-fixed) equations of motion 
\rf{gamma-p-0}--\rf{X-eqs-0} describing its dynamics in the extra dimensions 
reduce to:
\br
\ydot^\a\!\!\partder{}{y^\a}A \bv_{y=y(\t)} = 0 \quad ,\quad
\ydot^\a\!\!\partder{}{y^\a}C \bv_{y=y(\t)}=0 \; ,
\lab{A-C-rel} \\
- A(y(\t)) + B_{\a\b}(y(\t)) \ydot^\a \ydot^\b = 0 \; ,
\lab{gamma-eqs-2} \\
\yddot^{\,\a} + \ydot^\b \ydot^\g \G^\a_{\b\g} + 
B^{\a\b} \(\frac{p\, a_0}{C(y)}\partder{}{y^\b} C(y) + 
\h\partder{}{y^\b} A(y)\)\bv_{y=y(\t)} = 0
\lab{y-eqs-1}
\er
(recall $a_0 = \mathrm{const}$ as in \rf{gamma-eqs-u}).

\textbf{Example 1: Two Flat Extra Dimensions.} In this case:
\br
y^\a = (\rho,\phi) \quad ,\quad 
B_{\a\b}(y) dy^\a dy^\b = d\rho^2 + \rho^2 d\phi^2 \; ;
\lab{flat} \\
A= A(\rho) \;\; ,\;\; C= C(\rho) \quad ;\;\; \rho = \rho_0 = \mathrm{const}
\;\; ; \;\; \phi (\t) = \om \t \; ,
\lab{A-C-rotation-flat}
\er
where:
\be
\om^2 = \frac{A(\rho_0)}{\rho_0^2} \quad ,\quad
A(\rho_0) = \rho_0 
\(\frac{p\, a_0}{C(\rho)} \pa_\rho C + \h \pa_\rho A)\)\bv_{\rho=\rho_0} \; .
\lab{orbit-flat}
\ee
Thus, we find that the \textsl{LL-brane} performs a {\em planar circular 
motion} in the flat extra dimensions whose radius $\rho_0$ and angular velocity 
$\om$ are determined from \rf{orbit-flat}. This property of the \textsl{LL-branes}
has to be contrasted with the usual case of Nambu-Goto-type braneworlds which
(in the ground state) occupy a {\em fixed position} in the extra dimensions.\\

\textbf{Example 2: Toroidal Extra Dimensions.} In this case:
\be
y^\a = (\th,\phi)\;\; ,\;\; 0 \leq \th,\phi \leq 2\pi \quad ,\quad 
B_{\a\b}(y) dy^\a dy^\b = d\th^2 + a^2 d\phi^2
\lab{toroidal}
\ee
The solutions read:
\be
\th (\t) = \om_1 \t \quad ,\quad \phi (\t) = \om_2 \t
\lab{y-eqs-toroidal-1}
\ee
where the admissible form of the background metric must be:
\be
A = A (\th - N\phi)\;\; ,\;\; C = C (\th - N\phi) \quad ,\quad
A^\pr (0) = 0 \;\; ,\;\; C^\pr (0) = 0 \; ,
\lab{A-C-toroidal-3}
\ee
($N$ -- arbitrary integer), with angular frequencies $\om_{1,2}$ in 
\rf{y-eqs-toroidal-1}:
\be
(\om_1)^2 = \frac{A(0)}{1 + a^2/N^2} \quad , \quad \om_2 = \frac{\om_1}{N} \; .
\lab{omega-eq}
\ee
We conclude that the LL-brane performs a {\em spiral motion} in the 
toroidal extra dimensions with winding frequencies as in \rf{omega-eq}.

\section{Lightlike Branes in Gravitational Backgrounds:\\ Codimension-One}
This case is qualitatively different from the case of codimension\\ $\geq 2$.
Here the metric \rf{nonstatic-metric} acquires the form of a general 
spherically symmetric metric:
\be
ds^2 = - A(t,y)(dt)^2 + C(t,y) h_{ij}(\vec{\th}) d\th^i d\th^j + B (t,y) (dy)^2
\lab{spherical-metric}
\ee
where $\vec{x} \equiv \vec{\th}$ are the angular coordinates
parametrizing the sphere $S^p$. 

The \textsl{LL-brane} equations of motion \rf{gamma-p-0}--\rf{X-eqs-0} now take 
the form: 
\br
-A + B \ydot^2 = 0 \;\; ,\; \mathrm{i.e.}\;\; \ydot = \pm \sqrt{\frac{A}{B}}
\quad ,\quad
\pa_t C + \ydot \pa_y C = 0
\lab{r-const} \\
\pa_\t \chi + \chi \llb \pa_t \ln \sqrt{AB} 
\pm \frac{1}{\sqrt{AB}} \Bigl(\pa_y A + p\, a_0 \pa_y \ln C\Bigr)\rrb_{y=y(\t)} = 0
\lab{X0-eq-1}
\er

First let us consider static spherically symmetric metrics in standard coordinates:
\be
ds^2 = - A(y)(dt)^2 + A^{-1}(y) (dy)^2 + y^2 h_{ij}(\vec{\th}) d\th^i d\th^j
\lab{standard-spherical}
\ee
where $y\equiv r$ is the radial-like coordinate. Here we obtain:
\be
\ydot = 0 \;\; ,\;\; \mathrm{i.e.}\;\; y(\t) = y_0 = \mathrm{const} \quad, \quad
A(y_0) = 0 \; ,
\lab{horizon-standard}
\ee
implying that the \textsl{LL-brane} positions itself {\em automatically} on the
horizon $y_0$ of the background metric (``horizon straddling''). Further, for
the dynamical tension we get:
\be
\chi (\t) = \chi_0 
\exp\lcurl\mp \t \(\pa_y A\bv_{y=y_0} + \frac{2 p\, a_0}{y_0}\)\rcurl
\quad ,\quad \chi_0 = \mathrm{const} \; .
\lab{chi-eq-standard-sol}
\ee
Thus, we find a time-asymmetric solution for the dynamical
brane tension which (upon appropriate choice of the signs $(\mp)$ depending on the 
sign of the constant factor in the exponent on the r.h.s. of \rf{chi-eq-standard-sol})
{\em exponentially ``inflates'' or ``deflates''} for large times.

Next consider spherically symmetric metrics in Kruskal-Szekeres-like coordinates:
\be
ds^2 = A\(y^2 - t^2\) \llb -(dt)^2 + (dy)^2\rrb + 
C\(y^2 - t^2\)\, h_{ij}(\vec{\th}) d\th^i d\th^j
\lab{kruskal-like}
\ee
where $(t,y)$ play the role of Kruskal-Szekeres's $(v,u)$ coordinates for 
Schwarzschild metrics \ct{kruskal}. Here the \textsl{LL-brane} equations of motion 
yield:
\be
\ydot = \pm 1 \;\;, \;\; \mathrm{i.e.}\;\; y(\t) = \pm \t \quad ,\quad
\(y^2 - t^2\)\bv_{t=\t,y=y(\t)} = 0 \; ,
\lab{horizon-kruskal}
\ee
\textsl{i.e.}, again the \textsl{LL-brane} locates itself {\em automatically} on the
horizon (``horizon straddling''), whereas for the dynamical tension we obtain:
\be
\chi (\t) = \chi_0 \exp\lcurl -\t^2\,\frac{p\, a_0\, C^{\pr}(0)}{A(0)C(0)}\rcurl \; .
\lab{chi-eq-kruskal-sol}
\ee
Thus, we find a time-symmetric {\em ``inflationary'' or ``deflationary''} solution 
with {\em quadratic} time dependence in the exponential for the dynamical brane 
tension (depending on the sign of the constant factor in the exponent on the 
r.h.s. of \rf{chi-eq-kruskal-sol}).

Let us also consider ``cosmological''-type metrics:
\be
ds^2 = - (dt)^2 + S^2(t)\llb (dy)^2 + f^2 (y) h_{ij}(\vec{\th}) d\th^i d\th^j\rrb \,
\lab{FRW-metric}
\ee
where $f(y)= y,\, \sin (y),\, \sinh (y)$. The \textsl{LL-brane} equations of 
motion give:
\be
\ydot = \pm \frac{1}{S(\t)} \quad ,\quad
S^2(\t)\, f^2(y(\t)) = \frac{1}{c_0^2} \;\; ,\;\;
c_0 = \textrm{const} \; ,
\lab{horizon-cosmolog}
\ee
implying: $S(\t) = \pm \frac{1}{c_0\, y_0}\, e^{-c_0\,\t}~$,
$S(\t) = \pm \frac{1}{c_0}\,\cosh\(c_0 (\t+\t_0)\)$ or\\
$S(\t) = \mp \frac{1}{c_0}\,\sinh\(c_0 (\t+\t_0)\)$, respectively,
where $y_0,\t_0 = \mathrm{const}$.

For the dynamical brane tension we obtain ``inflation''/``deflation'' at 
$\t \to \pm\infty$:
\be
\chi (\t) = \chi_0 \bigl( S(\t)\bigr)^{2p\, a_0-1} \;\; ,\;\; 
\chi_0 = \mathrm{const}
\lab{chi-eq-cosmolog}
\ee

\textbf{Example 1: de Sitter embedding space metric in Kruskal-Szekeres-like
(Gibbons-Hawking \ct{gibbons-hawking}) coordinates}. In this case:
\br
ds^2 = A(y^2-t^2)\llb - (dt)^2 + (dy)^2\rrb +
R^2 (y^2-t^2) h_{ij}(\vec{\th}) d\th^i d\th^j
\lab{gibbons-hawking-metric} \\
A(y^2-t^2) = \frac{4}{K(1+y^2-t^2)^2} \quad ,\;\;
R (y^2-t^2) = \frac{1}{\sqrt{K}}\,\frac{1-(y^2-t^2)}{1+y^2-t^2}
\lab{gibbons-hawking-metric-1}
\er
($K$ is the cosmological constant).\\
We obtain exponential ``inflation'' at $\t \to \pm \infty$
for the dynamical tension of \textsl{LL-branes} occupying de Sitter horizon:
\be
\chi(\t) = \chi_0 \exp\bigl\{\t^2\,p\,a_0 K\bigr\} \;.
\lab{chi-gibbons-hawking}
\ee

\textbf{Example 2: Schwarzschild background metric in Kruskal-Szekeres coordinates
\ct{kruskal}}.
In this case (here we take $D=p+2=4$)
\be
ds^2 = A(y^2-t^2)\llb - (dt)^2 + (dy)^2\rrb +
R^2 (y^2-t^2) h_{ij}(\vec{\th}) d\th^i d\th^j
\lab{kruskal-metric}
\ee
\be
A = \frac{4R_0^3}{R} \exp\lcurl - \frac{R}{R_0}\rcurl \quad ,\quad
\(\frac{R}{R_0}-1\) \exp\lcurl\frac{R}{R_0}\rcurl = y^2 - t^2
\lab{kruskal-metric-1}
\ee
(here $R_0 \equiv 2G_N m$). We obtain exponential ``deflation'' at 
$\t \to \pm \infty$ for the dynamical tension of \textsl{LL-branes} sitting on 
the Schwarzschild horizon:
\be
\chi(\t) = \chi_0 \exp\lcurl-\t^2\,\frac{a_0}{R^2_0}\rcurl \; .
\lab{chi-kruskal}
\ee

\textbf{Example 3: Reissner-Nordstr{\"om} background metric in Kruskal-Szekeress-like 
coordinates}. In this case (here again $D=p+2=4$):
\be
ds^2 = A (y^2-t^2)\llb - (dt)^2 + (dy)^2\rrb +
R^2 (y^2-t^2) g_{ij}(\vec{\th}) d\th^i d\th^j \; .
\lab{RN-metric}
\ee
In the region around the outer Reissner-Nordstr{\"om} horizon $R=R_{(+)}$, 
\textsl{i.e.}, for $R>R_{(-)}$ ($R=R_{(-)}$ -- inner Reissner-Nordstr{\"om}
horizon), the functions $A(x),\, R(x)$ are defined as:
\br
y^2 -t^2 = \frac{R-R_{(+)}}{\( R-R_{(-)}\)^{R^2_{(-)}/R^2_{(+)}}}\,
\exp\lcurl R \frac{R_{(+)}-R_{(-)}}{R^2_{(+)}} \rcurl
\lab{RN-kruskal-outer-1} \\
A (y^2-t^2) = 
\frac{4 R^4_{(+)} \( R-R_{(-)}\)^{1+R^2_{(-)}/R^2_{(+)}}}{\( R_{(+)}-R_{(-)}\)^2 R^2}\,
\exp\lcurl - R \frac{R_{(+)}-R_{(-)}}{R^2_{(+)}} \rcurl
\lab{RN-kruskal-outer-2}
\er
We find here {\em exponentially ``deflating''} tension for the
\textsl{LL-brane} sitting on the outer Reissner-Nordstr{\"om} horizon:
\be
\chi(\t) = \chi_0 \exp\lcurl -\t^2\,\frac{a_0}{R^2_{(+)}}
\( 1 - \frac{R_{(-)}}{R_{(+)}}\) \rcurl
\lab{chi-RN-outer-1}
\ee
(a phenomenon similar to the case of LL-brane sitting on Schwarzschild horizon 
\rf{chi-kruskal}).

In the region around the inner Reissner-Nordstr{\"om} horizon $R=R_{(-)}$, 
\textsl{i.e.}, for $R<R_{(+)}$, the functions $A(x),\, R(x)$ are given by:
\br
y^2 -t^2 = \frac{R-R_{(-)}}{\( R-R_{(+)}\)^{R^2_{(+)}/R^2_{(-)}}}\,
\exp\lcurl R \frac{R_{(-)}-R_{(+)}}{R^2_{(-)}} \rcurl
\lab{RN-kruskal-inner-1} \\
A (y^2-t^2) = 
\frac{4 R^4_{(-)} \( R_{(+)}-R\)^{1+R^2_{(+)}/R^2_{(-)}}}{\( R_{(-)}-R_{(+)}\)^2 R^2}\,
\exp\lcurl - R \frac{R_{(-)}-R_{(+)}}{R^2_{(-)}} \rcurl
\lab{RN-kruskal-inner-2}
\er
In this case we obtain {\em exponentially ``inflating''} tension for the
\textsl{LL-brane} occupying the inner Reissner-Nordstr{\"om} horizon:
\be
\chi(\t) = \chi_0 \exp\lcurl \t^2\,\frac{a_0}{R^2_{(-)}}
\(\frac{R_{(+)}}{R_{(-)}} - 1\) \rcurl \; .
\lab{chi-RN-outer-2}
\ee
The latter effect is similar to \rf{chi-gibbons-hawking} --
the exponential brane tension ``inflation'' on de Sitter horizon.

\section{Lightlike Branes in Kerr Black Hole and Black String  Backgrounds}
Let us consider $D\! =\! 4$-dimensional Kerr background metric in the 
standard Boyer-Lindquist coordinates (see \textsl{e.g.} \ct{carroll}):
\be
ds^2 = -A (dt)^2 - 2 E dt\,d\vp + \frac{\S}{\D} (dr)^2 + \S (d\th)^2 +
D \sin^2 \th (d\vp)^2)  \; ,
\lab{Kerr-metric}
\ee
\br
A\equiv \frac{\D - a^2 \sin^2 \th}{\S} \;\; ,\;\;
E\equiv \frac{a \sin^2 \th\,\(r^2 + a^2 - \D\)}{\S} \\
\nonu \\
D\equiv \frac{\( r^2 + a^2\)^2 - \D a^2 \sin^2 \th}{\S} \; ,
\phantom{aaaaaaa}
\lab{Kerr-coeff}
\er
where $\S \equiv r^2 + a^2\cos^2 \th\; ,\; \D \equiv r^2 + a^2 - 2 Mr$, and
the following ansatz for the \textsl{LL-brane} embedding (here $p=2$):
\be
X^0 \equiv t = \t \;\; ,\;\; r=r(\t) \;\; ,\;\; \th = \s^1 \;\; ,\;\;
\vp = \s^2 + {\wti \vp}(\t) \; .
\lab{Kerr-ansatz}
\ee
In this case the \textsl{LL-brane} equations of motion
\rf{gamma-p-0}--\rf{gamma-eqs-0} acquire the form:
\br
-A + \frac{\S}{\D} \rdot^2 + D \sin^2 \th\,\vpdot^2 - 2 E \vpdot = 0
\nonu \\
-E + D \sin^2 \th\,\vpdot = 0 \quad ,\quad  
\frac{d}{d\t}\( D\S \sin^2 \th\) = 0 \; .
\lab{gamma-eqs-kerr}
\er
Inserting the ansatz \rf{Kerr-ansatz} into \rf{gamma-eqs-kerr} we obtain:
\br
r = r_0 = \mathrm{const}\;\; \mathrm{with}\;\; \D (r_0) = 0 \quad ,\quad
i.e. \; r_0 \; - \; \mathrm{Kerr ~horizon}\; ,
\lab{kerr-straddle}\\
\om \equiv \vpdot = \frac{a}{r_0^2 + a^2} \; - \; 
\mathrm{constant ~angular ~velocity} \; .
\lab{kerr-dragging}
\er
Among the $X^\m$-equations of motion \rf{X-eqs-0} only the $X^0$-equation
yields additional information, namely, we obtain from the latter an exponential 
{\em ``inflating''/''deflating''} solution for the dynamical
\textsl{LL-brane} tension in Kerr black hole background:
\be
\chi (\t) = \chi_0 \,\exp\Bigl\{\mp\t\,\Bigl(\frac{1}{M}-\frac{1}{r_0}\Bigr)\Bigr\}
\; .
\lab{chi-eq-kerr}
\ee

From \rf{kerr-straddle}--\rf{chi-eq-kerr} we conclude that, similarly to the
spherically symmetric case, \textsl{LL-branes} moving as test branes in Kerr
rotating black hole background automatically straddle the Kerr horizon and
in addition they are ``dragged'' (rotate along) with angular velocity
$\om$ given in \rf{kerr-dragging}. Note that the latter expression coincides
precisely with the definition of Kerr horizon's angular velocity (Eq.(6.92) in
ref.\ct{carroll}). Furthermore, as in the spherically symmetric case we find 
``mass inflation/deflation'' effect on Kerr horizon via the exponential time
dependence of the dynamical \textsl{LL-brane} tension. 

The above analysis applies straightforwardly to the case of lightlike string
($p=1$) moving in $D=4$ Kerr black hole background, \textsl{i.e.}, a case of 
codimension two. Here the lightlike string positions itself automatically on the 
equator of the horizon \rf{kerr-straddle} ($r=r_0\; ,\; \th = \frac{\pi}{2}$) and 
again rotates along the latter with the angular velocity \rf{kerr-dragging}.

Along the same lines we can analyze the dynamics of codimension-two test
\textsl{LL-brane} ($p=2$) in a $D=5$ boosted black string background
\ct{boosted-BS}. The metric of the latter reads:
\br
ds^2 =  - A (dt)^2 + \frac{(dr)^2}{f} + r^2 \llb (d\th)^2 +
\sin^2 \th\,(d\vp)^2\rrb + B (dz)^2 - 2\cE dt\,dz
\nonu\\
\phantom{aaaaaa}
\lab{boosted-BS}
\er
\br
A(r) \equiv 1 - (1-f(r))\,\cosh^2\b \quad ,\quad
B(r) \equiv 1 + (1-f(r))\,\sinh^2\b \; ,
\nonu \\
\cE (r) \equiv - (1-f(r))\,\sinh\b\,\cosh\b \quad ,\quad
f(r) \equiv 1 - \frac{r_0}{r} \; ,\phantom{aaaaaa}
\lab{boosted-BS-coeff}
\er
where $\b$ is the boost rapidity parameter,
and we employ the following ansatz for the \textsl{LL-brane} embedding:
\be
X^0 \equiv t = \t \;\; ,\;\; r=r(\t) \;\; ,\;\; \th = \s^1 \;\; ,\;\;
\vp = \s^2 \;\; ,\;\; z = z(\t) \; .
\lab{BS-ansatz}
\ee
Inserting \rf{BS-ansatz} into \textsl{LL-brane} equations of motion 
\rf{gamma-p-0}--\rf{X-eqs-0} we obtain:
\be
r(\t) = r_0 \quad ,\quad z(\t) = \om\,\t + z_0 \;\;\; \mathrm{with}\;\;
\om = - \tanh\b \quad ,\quad \chi = \mathrm{const} \; .
\lab{BS-sol}
\ee
In other words the codimension-two test \textsl{LL-brane} automatically
occupies the sphere $S^2$ of the $S^2\times S^1$ horizon of the boosted black
string and winds the circle $S^1$ of the horizon with angular velocity
$\om$ given in \rf{BS-sol}.

\section{Further Developments and Outlook}
Codimension one \textsl{LL-branes} possess natural couplings to bulk Maxwell 
$\cA_\m$ and Kalb-Ramond $\cA_{\m_1 \ldots \m_{p+1}}$ gauge fields ($D-1=p+1$,
see refs.\ct{will-brane-all}):
\br
\wti{S}_{\mathrm{LL}} = \int d^{p+1}\s \,\P (\vp)
\llb - \h \g^{ab} \pa_a X^{\m} \pa_b X^{\n} G_{\m\n} (X) + L\!\( F^2\)\rrb
\nonu \\
-q \int d^{p+1}\s \vareps^{ab_1\ldots b_p} F_{b_1\ldots b_p} \pa_a X^\m \cA_\m (X)
\nonu \\
-\frac{\b}{(p+1)!} \int d^{p+1}\s \vareps^{a_1\ldots a_{p+1}}
\pa_{a_1} X^{\m_1} \ldots \pa_{a_{p+1}} X^{\m_{p+1}}\cA_{\m_1\ldots\m_{p+1}}(X)
\lab{LL-brane-ext}
\nonu
\er
As shown in \ct{will-brane-all} by considering bulk 
Einstein-Maxwell+Kalb-Ramond-field system coupled to a \textsl{LL-brane}:
\br
S = \int\!\! d^D x\,\sqrt{-G}\,\llb \frac{R(G)}{16\pi G_N}
- \frac{1}{4e^2} \cF_{\m\n}\cF^{\m\n} 
- \frac{1}{p! 2} \cF_{\m_1 \ldots\m_D}\cF^{\m_1 \ldots\m_D}\rrb
+ \wti{S}_{\mathrm{LL}} \phantom{aaaa}
\lab{E-M-WILL}
\er
the \textsl{LL-brane} can serve as a material and charge source for gravity 
and electromagnetism and, furthermore, it generates {\em dynamical} cosmological 
constant through the coupling to the Kalb-Ramond bulk field:
\be
K = \frac{8\pi G_N}{p(p+1)} \b^2  \; .
\lab{dyn-cosmolog-const}
\ee
There exist the following static spherically symmetric solutions of the
coupled system. The bulk space-time consists of two regions with different
geometries separated by a common horizon occupied by the \textsl{LL-brane}. 
The matching of the metric components across the horizon reads (using the
same notations as in \rf{standard-spherical}):
\br
A_{(-)}(y_0) = 0 = A_{(+)}(y_0) \;\;, \;\;
\(\pa_y A_{(+)} - \pa_y A_{(-)}\)_{y=y_0} = 
- \frac{16\pi G_N}{(2a_0)^{p/2 -1}}\,\chi \phantom{aa}
\lab{matching}
\er
As discussed in more details in a forthcoming paper \ct{varna-08}, conditions
\rf{matching} allow for a {\em non-singular} black hole type solution where
the geometry of the interior region (below the horizon) 
is de-Sitter with {\em dynamically} generated cosmological constant $K$
\rf{dyn-cosmolog-const}, whereas the outer region's geometry (above the
horizon) is Schwarzschild or Reissner-Nordstr{\"o}m.

\vspace{.1in}
\textbf{Acknowledgements.}
{\small E.N. and S.P. are sincerely grateful to Prof. Branko Dragovich and the organizers
of the Fifth Summer School in Modern Mathematical Physics (Belgrade, June 2008) 
for cordial hospitality. 
E.N. and S.P. are supported by European RTN network {\em ``Forces-Universe''} 
(contract No.\textsl{MRTN-CT-2004-005104}).
They also received partial support from Bulgarian NSF grant \textsl{F-1412/04}.
Finally, all of us acknowledge support of our collaboration through the exchange
agreement between the Ben-Gurion University of the Negev (Beer-Sheva, Israel) and
the Bulgarian Academy of Sciences.}

\end{document}